\documentclass[journal=jpccck,manuscript=article]{achemso}
\setkeys{acs}{articletitle=true}
\usepackage{chemformula} 
\usepackage[T1]{fontenc} 
\usepackage{amssymb}
\usepackage{amsmath}

\usepackage{multirow}
\usepackage{booktabs}
\usepackage{subcaption}

\author{Magdalena Laurien}
\email{laurienm@mcmaster.ca}
\affiliation{Department of Materials Science and Engineering, McMaster University, 1280 Main Street West,
Hamilton, Ontario L8S 4L7, Canada}
\author{Himanshu Saini}
\affiliation{Department of Materials Science and Engineering, McMaster University, 1280 Main Street West,
Hamilton, Ontario L8S 4L7, Canada}
\author{Oleg Rubel}
\email{rubelo@mcmaster.ca}
\affiliation{Department of Materials Science and Engineering, McMaster University, 1280 Main Street West,
Hamilton, Ontario L8S 4L7, Canada}

\title[An \textsf{achemso} demo]
  {Band alignment of monolayer \ch{CaP3}, \ch{CaAs3}, \ch{BaAs3} and the role of $p$-$d$ orbital interactions in the formation of conduction band minima} 

\keywords{density functional theory}

\RequirePackage[normalem]{ulem}
\RequirePackage{color}\definecolor{RED}{rgb}{1,0,0}\definecolor{BLUE}{rgb}{0,0,1}

\begin{document}

\begin{tocentry}

\includegraphics{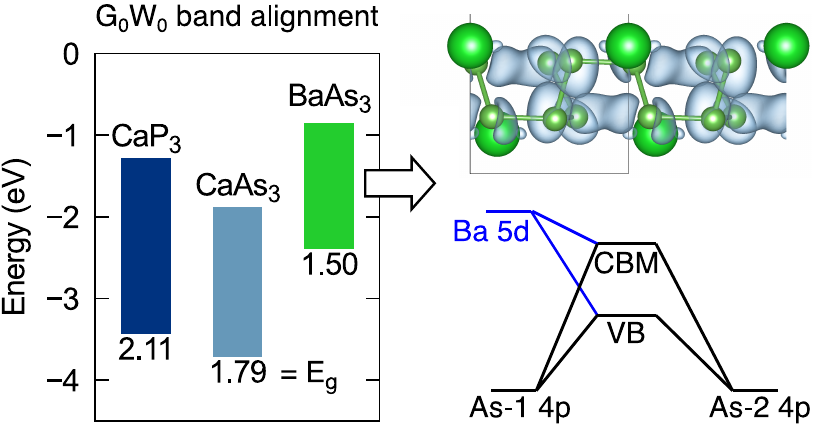}

\end{tocentry}

\begin{abstract}
Recently, a number of new two-dimensional (2D) materials based on puckered phosphorene and arsenene have been predicted with moderate band gaps, good absorption properties and carrier mobilities superior to transition metal dichalcogenides. For heterojunction applications, it is important to know the relative band alignment of these new 2D materials. We report the band alignment of puckered \ch{CaP3}, \ch{CaAs3} and \ch{BaAs3} monolayers at the quasiparticle level of theory (G$_0$W$_0$), calculating band offsets for isolated monolayers according to the electron affinity rule. Our calculations suggest that monolayer \ch{CaP3}, \ch{CaAs3} and \ch{BaAs3} all form type-II (staggered) heterojunctions. Their quasiparticle gaps are 2.1 (direct), 1.8 (direct) and 1.5~eV (indirect), respectively.
We also examine trends in the electronic structure in the light of chemical bonding analysis. We show that the indirect band gap in monolayer \ch{BaAs3} is caused by relatively strong As~$3p$ - Ba~$5d$ bonding interactions that stabilize the conduction band away from the $\Gamma$ point between $\Gamma$ and $S$.
\end{abstract}

\section{Introduction}

The optical and electronic properties of black phosphorus (bP), such as its high carrier mobility~\citep{li_black_2014} and tunable band gap~\citep{tran_layer-controlled_2014}, have sparked interest in the broader family of phosphorene related 2D material structures. Recently, several new 2D materials related to phosphorene and arsenene -- for which a part of the As or P atoms is replaced by group II elements~\citep{tang_baas3_2019, li_tunable_2019, lu_cap3_2018} or elements of other groups (III, IV, V) \citep{jing_gep3_2017, jalil_phosphorene-like_2020, miao_tunable_2017, liu_two-dimensional_2020} -- have been predicted. These layered materials show a combination of high carrier mobilities, moderate band gaps and good light absorption properties promising for next-generation electronic heterojunction devices like solar cells and transistors.
 
For example, monolayer (1L) \ch{CaP3} is predicted to have a band gap of 1.15~eV \citep{lu_cap3_2018} (using a hybrid functional, HSE06) and a high electron mobility of up to ca. 20~000~cm$^2$~V$^{-1}$~s$^{-1}$ (using deformation potential theory, value needs to be taken with caution as deformation potential theory usually overestimates mobility values~\citep{gaddemane_theoretical_2018, cheng_why_2020}).

For materials selection in heterojunction design, it is necessary to know the band alignments of the layered structures. However, band alignments for novel 2D materials of the phosphorene family have not yet been reported and the majority of the band gaps of these 2D structures has been calculated using hybrid functionals at the density functional level of theory (DFT). Although hybrid functionals alleviate the band gap problem of DFT, it is generally accepted that GW quasiparticle calculations provide the most reliable and accurate level of theory for band structure and band alignment calculations.\citep{garza_predicting_2016, huser_quasiparticle_2013, jiang_electronic_2012, koda_trends_2018}

Here, we calculate the band gaps of the recently predicted\citep{lu_cap3_2018, li_tunable_2019, tang_baas3_2019} \ch{CaP3}, \ch{CaAs3} and \ch{BaAs3} monolayers at the G$_0$W$_0$ level of theory. We estimate their band alignments according to the electron affinity rule which states that the conduction band offset of a heterojunction can be obtained from the electron affinities of the individual materials. Analyzing the electron affinity of the isolated monolayers is a valid first step in materials selection process for possible 2D heterostructures. \citep{guo_band_2016, chiu_band_2017} Further, we explain the differences in band alignment based on structure analysis and show that the indirect nature of the band gap in 1L \ch{BaAs3} is caused by increasing $p$-$d$ orbital interactions. 

\section{Methods}
\subsection{Structural relaxation (DFT)}

First, we relaxed the experimental bulk structures of CaAs$_{3}$, \citep{bauhofer_structure_1981} CaP$_{3}$, \citep{dahlmann_cap3_1973} and BaAs$_{3}$ \citep{bauhofer_structure_1981}. The structures were relaxed on the DFT level using the Vienna ab initio package (VASP) \citep{kresse_efficient_1996, kresse_efficiency_1996} with projector-augmented potentials\citep{blochl_projector_1994} as implemented by Kresse and Joubert \citep{kresse_ultrasoft_1999}. We explicitly included eight valence and core electrons for Ca, ten electrons for Ba, 15 electrons for As and five valence electrons for P. The Perdew-Burke-Ernzerhof (PBE) \citep{perdew_generalized_1996} exchange-correlation functional with D3 van-der-Waals corrections\citep{grimme_consistent_2010} was used to obtain the electronic ground state. We used a $k$ grid sampling of $6 \times 6 \times 6$ and a plane-wave-expansion cutoff of 550~eV. The structures were fully relaxed until the residual atomic forces were less than 0.001~eV/{\AA}. From these, monolayer structures were obtained and subsequently relaxed with VASP. We used a $6 \times 6 \times 1$ $k$ grid and fixed the out-of-plane lattice parameter $c$ during the relaxation procedure. To avoid interactions between periodic images, we included a vacuum of at least 22.5~{\AA} in the out-of-plane direction. The POSCAR structure files of the relaxed monolayers are included in the supplementary information. 

\subsection{G$_0$W$_0$ calculations}

As input for the GW calculation, we computed DFT wavefunctions and eigenvalues with the Quantum Espresso (QE) code \citep{giannozzi_quantum_2009}. QE band structure calculations were performed for the relaxed monolayer structures with PBE and norm-conserving pseudopotentials \citep{van_setten_pseudodojo_2018}, using a plane-wave energy cutoff of 116~Ry and $k$ grid of $12 \times 12 \times 1$. We included ten valence and semicore states for Ca and Ba, 15 states for As and five valence states for P. The vacuum included for the QE and subsequent calculations was smaller than during the structural relaxation and was sufficient to contain at least 99\% of the charge density in half the unit cell. We neglected spin orbit coupling in our calculations as it was shown by \citet{tang_baas3_2019} that including relativistic effects changes the band gap by less than 0.01~eV in monolayer \ch{BaAs3}. 

Subsequently, we performed "single-shot" G$_0$W$_0$ calculations with the BerkeleyGW software \citep{hybertsen_electron_1986, rohlfing_electron-hole_2000, deslippe_berkeleygw_2012}. We employed the generalized plasmon-pole model \citep{hybertsen_electron_1986}, the static-remainder technique \citep{deslippe_coulomb-hole_2013}, and truncation of the long-range Coulomb interaction \citep{ismail-beigi_truncation_2006}. We used a $k$ point sampling of $6 \times 6 \times 1$ and a kinetic energy cutoff of 20~Ry. About 1000 bands were included for the calculation of the dielectric function as well as for the calculation of the self-energy. To speed up the convergence of the quasiparticle band gap with respect to $k$ grid sampling, we employed nonuniform-neck subsampling \citep{da_jornada_nonuniform_2017} with 10 radial subpoints. The convergence was tested for the \ch{CaAs3} monolayer as a representative material. The quasiparticle band gap at $\Gamma$ was converged within 0.05~eV with respect to calculations with a kinetic cutoff of 40~Ry and 3000 bands in the summations, and a calculation with a $12 \times 12 \times 1$ $k$ point grid plus 10 subpoints. 

To verify our approach, we also calculated the quasiparticle band gaps (not to be confused with the optical band gap where excitonic effects contribute a significant correction) of the parent structures, puckered phosphorene and arsenene, using the same parameters. We obtained a band gap of 2.02~eV for phosphorene and 1.84~eV for arsenene. The value of phosphorene is in excellent agreement with other theoretical (\citet{qiu_environmental_2017}: 2.08~eV, \citet{da_jornada_nonuniform_2017}: 2.05~eV, \citet{lu_lightmatter_2016}: 2.0~eV), and experimental studies (\citet{liang_electronic_2014} with scanning tunneling spectroscopy: 2.05 eV). For puckered arsenene, experimental quasiparticle gaps are not available and theoretical band gaps vary between 1.54~eV~\citep{shu_electronic_2018}, 1.58~eV~\citep{kecik_stability_2016} and 2.16~eV~\citep{niu_arsenene-based_2017}. The discrepancies may, among other factors, be attributed to structural differences, as experimental structures of both 1L and bulk puckered arsenene, which would serve as starting point for DFT calculations, are still lacking.

The G$_0$W$_0$ band structure was obtained from Wannier interpolation with the wannier90 package~\citep{mostofi_updated_2014}. A comparison of the directly calculated and interpolated PBE band structure is shown in the supplementary information, Figs.~S1-S3, to illustrate the accuracy of the interpolation. The wannierization was performed for 100 iterations using random projections with 40 bands (20 valence and 20 conduction bands, each) as input to obtain 40 Wannier functions. 

\subsection{Data analysis}
Crystal structures were visualized with VESTA.~\citep{momma_vesta_2011} The Brillouin zone was visualized with XCrySDen.~\citep{kokalj_xcrysdennew_1999} The partial density of states (pDOS) and the projected charge density were obtained using VASP with the PBE exchange-correlation functional. 
The potential in vacuum was calculated at the DFT-PBE level using the vaspkit tool \citep{wang_vaspkit_2020}. We estimated the G$_0$W$_0$ band alignment (electron affinity) with respect to vacuum of each monolayer using the band-gap-center approximation \citep{liang_quasiparticle_2013}
\begin{equation}\label{Eq:electron-affinity}
	E_\text{EA} \approx
		\frac{1}{2}(E_\text{c}^\text{DFT-PBE} + E_\text{v}^\text{DFT-PBE}) + \frac{1}{2} E_\text{g}^{G_{0}W_{0}},
\end{equation}
which is equivalent to shifting the band edges of the PBE bands by $\pm (E_\text{g}^{G_{0}W_{0}}-E_\text{g}^\text{DFT-PBE})/2$. We note that the band-gap-center approximation works well for PBE centers and G$_0$W$_0$ gaps, however, it does not hold for  results obtained with hybrid functionals~\citep{wang_mos2_2014,slassi_interlayer_2020}.

To estimate the heterojunction type (e.g. straddled or staggered gap) between the isolated monolayers, we applied Anderson's electron affinity rule \citep{anderson_germanium-gallium_1960}. The electron affinity rule implies that the conduction band offset of a heterojunction can be obtained by aligning the vacuum levels of the two materials forming the junction. In other words, the band alignment is in the unpinned limit; no charge transfer across the junction is considered. It has been confirmed theoretically \citep{guo_band_2016} and experimentally \citep{chiu_band_2017} that the band alignment follows the electron affinity rule for vertically stacked heterostructures of transition metal dichalcogenides. This behaviour is attributed to the weak van der Waals bonding across the interface and the absence of dangling bonds in 2D materials \citep{guo_band_2016, chiu_band_2017}. Now, phosphorene and arsenene related materials are expected to have stronger interlayer interactions than transition metal dichalcogenides, which is indicated by the rapid decrease of the band gap with number of layers. Therefore, the heterojunction predictions from the electron affinity rule need to be taken with caution. With interlayer interactions turned on, the band offsets can change due to charge transfer across the layer and band mixing may occur~\citep{koda_trends_2018, koda_tuning_2017}. 

\section{Results and discussion}

\begin{figure}[h]
	\includegraphics{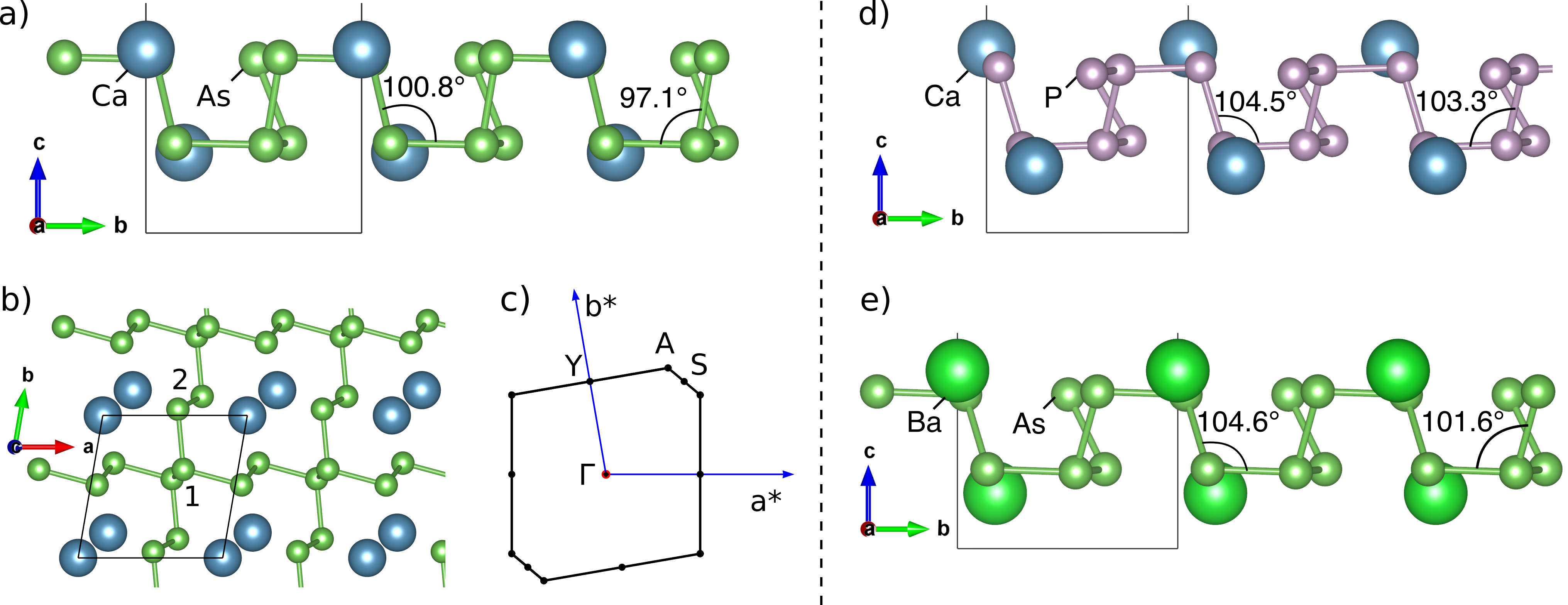}
	\caption{Structure of \ch{CaAs3}, \ch{CaP3} and \ch{BaAs3} monolayer. a) 1L \ch{CaAs3} as a representative compound shows a puckered configuration similar to arsenene where one quarter of As atoms is replaced with Ca, the other As atoms forming a polyanionic network. b) top view of the \ch{CaAs3} monolayer. The numbers 1 and 2 refer to two structural types of As atoms (see main text for details) c) Brillouin zone of the \ch{CaAs3} monolayer. The structures of monolayer \ch{CaP3} (d) and  \ch{BaAs3} (e)  are added to show the distinct structural distortion of each material.}
	\label{Fig:structure}
\end{figure}

\subsection{Structure and bonding}
Figure~\ref{Fig:structure}(a-c) shows the atomic structure and Brillouin zone of 1L \ch{CaAs3} as an example structure of the \ch{CaP3} family. The Ca atom replaces one quarter of the As atoms in the puckered structure of puckered arsenene. 1L \ch{CaP3} and 1L \ch{BaAs3} are also shown in Fig~\ref{Fig:structure}d) and e). 1L \ch{CaP3} and 1L \ch{CaAs3} inherit the $\overline{P}1$ space group from the corresponding bulk structures. 1L \ch{BaAs3} possesses higher symmetry and belongs to the $C2/m$ space group same as its bulk structure. 

In pristine arsenene, the As atoms form covalent bonds. For 1L \ch{CaAs3}, the Ca atom is stabilized by ionic interactions between the Ca cation and the anionic As mesh.~\citep{li_tunable_2019} We note that one can distinguish between two types of As atoms (see Fig~\ref{Fig:structure} b)): As1 which is further away from the Ca atoms, and As2 which is closer to the Ca atoms with As2 having a more negative partial charge than As1.~\citep{li_tunable_2019} (We note here that atoms belonging to each type are not necessarily symmetrically identical; they are only equivalent for \ch{BaAs3}). 
These considerations hold true in analogy for \ch{BaAs3} and \ch{CaP3}.

\begin{figure}[h]
	\includegraphics{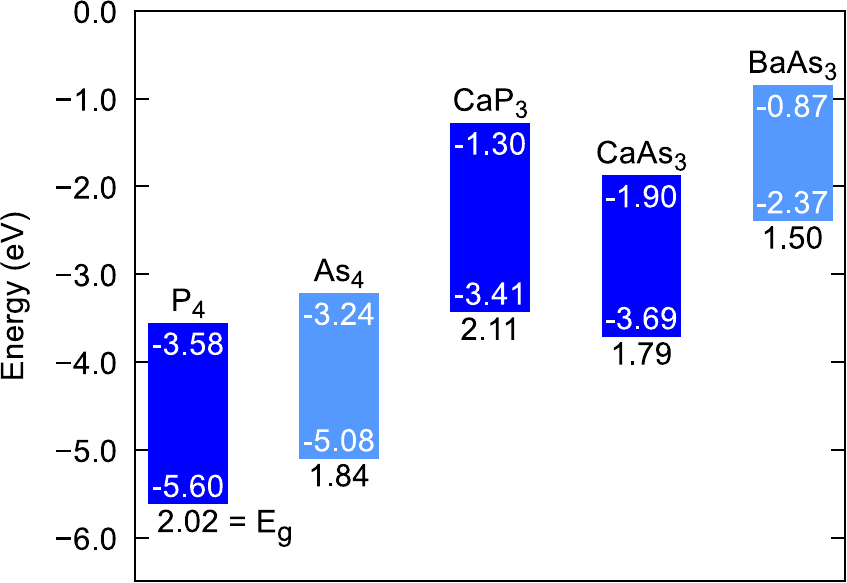}
	\caption{Band alignment of isolated monolayer \ch{CaP3}, \ch{CaAs3} and \ch{BaAs3} with respect to the vacuum level (0~eV). Band energies and offsets are obtained from G$_0$W$_0$ calculations with the band-gap center approximation (Eq.~(\ref{Eq:electron-affinity})). Band gaps and alignment of black phosphorene (P$_4$) and arsenene (As$_4$) have been added for reference. Indirect band gaps are indicated by light blue color, direct band gaps correspond to dark blue.}
	\label{Fig:alignment}
\end{figure}

\subsection{Band alignment}
The band alignments of monolayer \ch{CaP3}, \ch{CaAs3} and \ch{BaAs3} obtained from G$_0$W$_0$ calculations with the band-gap center approximation are shown in Fig~\ref{Fig:alignment}. The G$_0$W$_0$ gaps of 1L \ch{CaP3}, \ch{CaAs3} and \ch{BaAs3} are 2.11~eV, 1.79~eV and 1.50~eV, respectively. Phosphorene ("P$_4$") and arsenene ("As$_4$") have been added for comparison. Direct and indirect band gaps are signified by a dark and light blue color, respectively.
The band alignment diagram shows that, according to the electron affinity rule, 1L \ch{CaP3}, \ch{CaAs3} and \ch{BaAs3} form type-II (staggered) heterojunctions.  Between the pristine monolayers and the compound monolayers, 1L-bP/1L-\ch{CaP3} and 1L-bP/1L-\ch{BaAs3} combinations are predicted to have broken-gap band alignment (type III) whereas the other combinations show type-II alignments. Type-II heterojunctions are of interest for solar cell applications because they enable charge separation of photoinduced electrons and holes across the junction~\citep{jariwala_van_2017}. The narrow type-II and type-III alignments of the potential 1L-bP/1L-Ca$X_3$ and 1L-arsenene/1L-Ca$X_3$ ($X$~=~As,P) heterojunctions suggest promising material combinations for tunneling field effect transistor applications~\citep{lv_recent_2019}. Such devices would benefit from the high carrier mobilities predicted for the monolayers of the \ch{CaP3} family as well as from the very easily tunable band gap. We note that for real heterojunctions with enabled interlayer interactions, the band offsets may change due to the following effects: charge transfer across the interface, band mixing, and changes in the band gap~\citep{koda_trends_2018, koda_tuning_2017}. 

We will now analyze the calculated band alignments in further detail.
Looking at the band alignment (Fig.~\ref{Fig:alignment}), it becomes apparent that monolayer \ch{CaP3}, \ch{CaAs3} and \ch{BaAs3} have considerably smaller conduction band offsets with respect to the vacuum level than pure arsenene and phosphorene. Ca and Ba as group-II elements have much lower electron affinities than the group-V elements (As and P) thus leading to a higher energy of the conduction band edge in these compounds.
Between monolayer \ch{CaAs3} and \ch{BaAs3}, the midgap of \ch{CaAs3} is lower in energy than for \ch{BaAs3} (Fig.~\ref{Fig:alignment}). The chemical reason for this energy difference is not apparent at first sight. 

To differentiate between structural and chemical effects, we swapped the Ca atoms of the 1L \ch{CaAs3} structure with Ba atoms and vice versa \textit{without} a subsequent relaxation of atomic positions or lattice parameters. We found that the vacuum level offset changes to the opposite after swapping Ca and Ba atoms. The midgap energy of 1L \ch{BaAs3} in its original structure is lower in energy with respect to the vacuum level than the 1L \ch{CaAs3} structure with Ba atoms at the position of Ca. We thus conclude that the difference in band alignment can be attributed to the structural difference between the compounds rather than to the type of alkaline metal included in the puckered arsenic net. For example, the Ba atoms stick out of the monolayer plane more than the Ca atoms because of their larger atomic radius (compare Fig.~\ref{Fig:structure}a and Fig.~\ref{Fig:structure}e). Furthermore the puckered structure of 1L \ch{BaAs3} is stretched within the plane in comparison to 1L \ch{CaAs3} as the angles between the As atoms in the puckered mesh are greater for 1L \ch{BaAs3}. 

Interestingly, the band edges of 1L \ch{CaP3} are higher in energy than the ones of 1L \ch{CaAs3} although the midgap energy of phosphorene is lower than that of arsenene (see Fig~\ref{Fig:alignment}). This is likely caused by structural distortions similar to that of 1L \ch{BaAs3}. For example, the P-P bond length of 2.22~{\AA} on average in 1L \ch{CaP3} is much shorter than the As-As bond in 1L \ch{CaAs3} with 2.50~{\AA} and, as a result, the Ca atoms "stick out" of the monolayer much more than in 1L \ch{CaAs3} (see Fig~\ref{Fig:structure}~d). The puckered mesh also shows larger angles than in 1L \ch{CaAs3}.

\subsection{Electronic structure, the effect of $p$-$d$ interactions}
\begin{figure}[h]
	\includegraphics{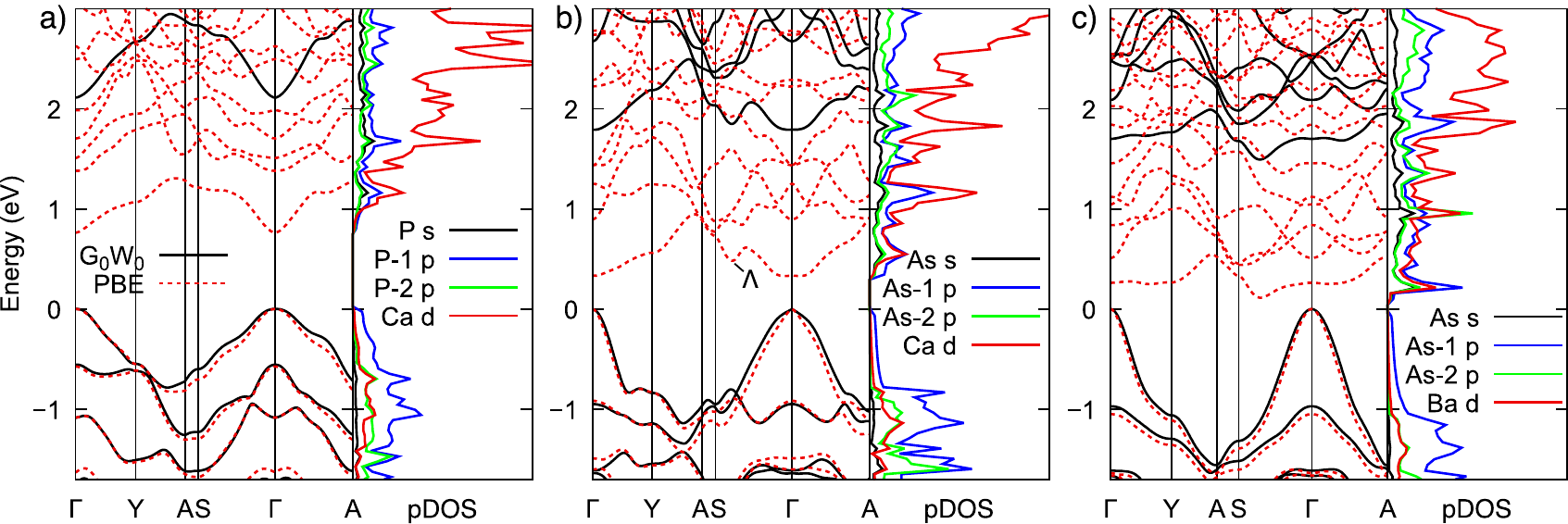}
	\caption{GW and PBE band structures from Wannier interpolation and projected density of states (pDOS) of a) monolayer \ch{CaP3}, b) monolayer \ch{CaAs3} and c) monolayer \ch{BaAs3}. The $\Lambda$ valley becomes the conduction band minimum for \ch{BaAs3}.}
	\label{Fig:band_DOS}
\end{figure}

We further explore and analyze the band structure and projected density of states (pDOS), pointing to trends from chemical bonding analysis. The PBE and G$_0$W$_0$ band structures of 1L \ch{CaP3}, 1L \ch{CaAs3} and 1L \ch{BaAs3} obtained by Wannier interpolation are shown in Fig~\ref{Fig:band_DOS}. 1L \ch{CaP3} and 1L \ch{CaAs3} have a direct band gap at $\Gamma$ while the band gap of 1L \ch{BaAs3} is indirect with the conduction band minimum (CBM) between S and $\Gamma$ . 
The valence band of the G$_0$W$_0$ band structure shows slightly less dispersion than that of the PBE band structure. The stronger dispersion of the PBE band structure is due to the underestimation of the band gap, which is consistent with the trend for other semiconductors~\citep{rubel_perturbation_2020} and can be rationalized in terms of a $k \cdot p$ theory that predicts $m^* \propto E_\text{g}$ scaling (see Ref.~\citep{yu_fundamentals_2010}, p.~71). Apart from the dispersion, the band gap correction is the only noticeable difference between the band structures obtained with PBE and G$_0$W$_0$. The pDOS shows that the valence band edge is almost entirely composed of As~$4p$ or P~$3p$ orbitals, with the type 2 atoms (adjacent to cations) making up the major part. The conduction band edge is again primarily composed of As~$4p$ (P~$3p$) states, As~$4s$ (P~$3s$) and metal $d$ states. 

We next discuss the indirect nature of the band gap in 1L \ch{BaAs3} versus the direct band gap in 1L \ch{CaAs3}. An interesting feature in the band structure of 1L \ch{CaAs3} is the local CBM between $S$ and $\Gamma$ at $k_x = 1/3, k_y = 1/3, k_z = 0$  which we shall call $\Lambda$ (see Fig~\ref{Fig:band_DOS}b). The same valley ($\Lambda$) is the global CBM for 1L \ch{BaAs3}. We want to shed light on why the $\Lambda$ valley is more stable with respect to the $\Gamma$ valley in 1L \ch{BaAs3} but not in 1L \ch{CaAs3}. 

\begin{figure}[!ht]
	\includegraphics{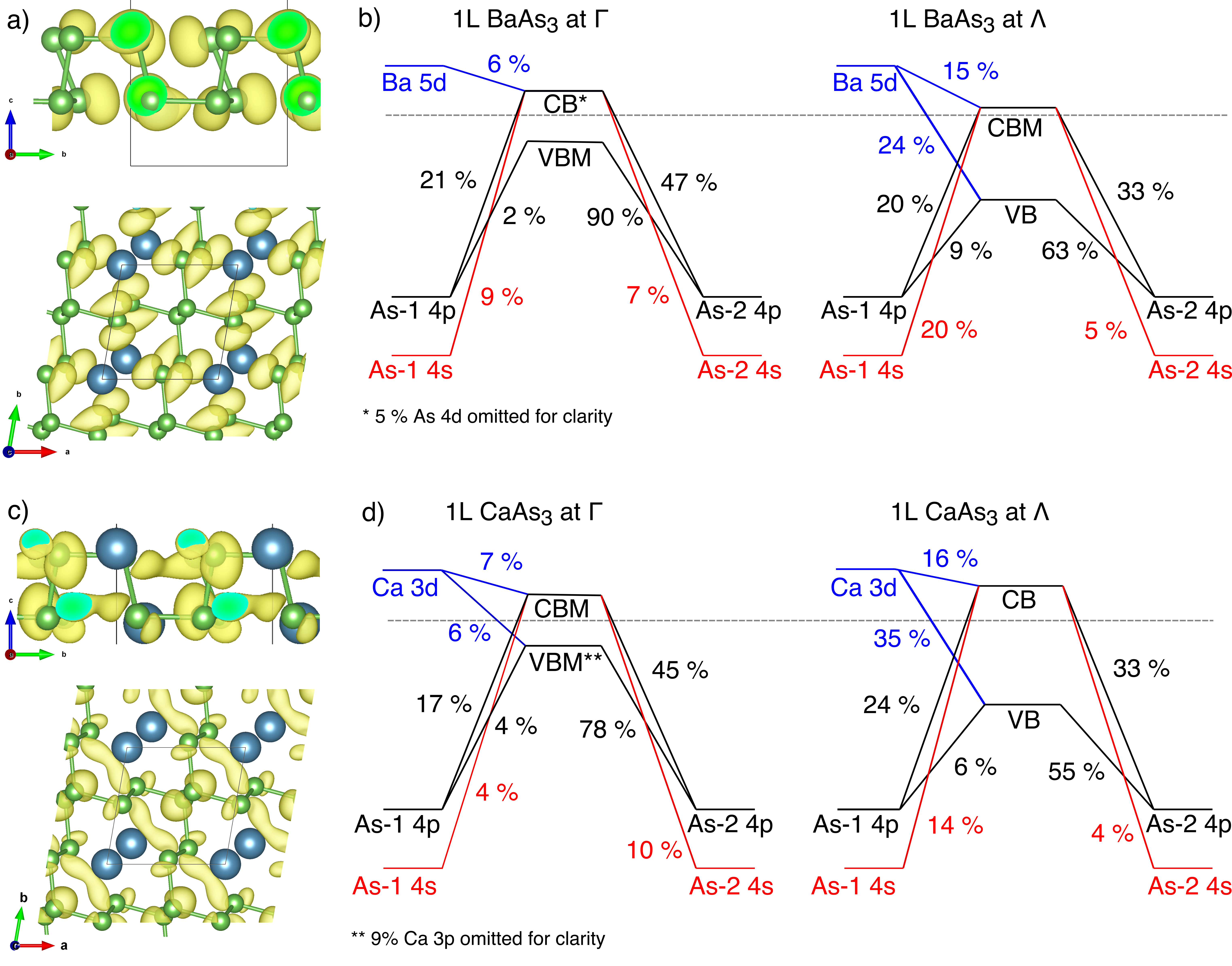}
	\caption{Orbital coupling in 1L \ch{CaAs3} and 1L \ch{BaAs3}. a) Band and $k$ point projected charge density of 1L \ch{CaAs3} for the valence band at the $\Gamma$ point. The isosurface is set at 0.004~e/\AA$^3$.  c) Band and $k$ point projected charge density for the conduction bands at the $\Lambda$ point. Isosurface: 0.004~e/\AA$^3$. b) and d) show schematics of the orbital coupling at the $\Gamma$ and $\Lambda$ points for 1L \ch{BaAs3} and 1L \ch{CaAs3}, respectively.}
	\label{Fig:charge_MOD}
\end{figure}

To that end, we analysed the orbital composition of the two valleys and then classified the interaction between the orbitals into bonding and antibonding by looking at the spatial distribution of the projected charge densities. From the pDOS we know that the valence band edge is almost entirely composed of As-2~$4p$ orbitals. Figure~\ref{Fig:charge_MOD}~a) shows the projected charge density of the topmost valence band at the $\Gamma$ point for 1L \ch{CaAs3}. The charge density distribution shows a clear $p$ character with lobes of adjacent atoms avoiding each other. This suggests that the $p$-$p$ interaction is antibonding. The conduction band edge (CBE) is mainly composed of As~$4p$, As~$4s$  and metal $d$ states. The projected charge density of the conduction band edge at $\Lambda$) for 1L \ch{CaAs3} is shown in Fig~\ref{Fig:charge_MOD}~c). The As-2~$4p$ - Ca~$3d$ orbital contributions are hybridized and appear as lobes of charge density in \ref{Fig:charge_MOD}~c) (see also Fig. S4 of the supplementary information). From the spatial distribution of the charge density we can infer that the hybridized As-2~$4p$ - Ca~$3d$ states at the CBE interact in a bonding fashion, while As-1 states are not involved. The same holds true for \ch{BaAs3}.

In addition to analyzing the projected charge density, crystal orbital Hamilton population (COHP) bonding analysis was performed with LOBSTER\citep{nelson_lobster_2020, deringer_crystal_2011, maintz_analytic_2013, dronskowski_crystal_1993}, which proved helpful to confirm the general picture of the antibonding nature at the valence and conduction band edges. However, the LOBSTER results are not shown because the current basis sets cannot project the empty $d$ states of Ca and Ba. The COHP analysis of the parent structures, arsenene and phosphorene, also clearly shows the antibonding nature of the $p$-$p$ interaction at the valence and conduction band edge (see supporting information, Fig.~S6-S7).

With the results of the bonding analysis we can create a schematic diagram of the coupling between As orbitals and Ca and Ba $d$ orbitals as shown in Figure~\ref{Fig:charge_MOD}~b), d). In comparison with the $\Gamma$ valley, the $\Lambda$ valley of the conduction band has higher contributions of As $s$ and metal $d$ orbitals. For example, the conduction band of 1L \ch{BaAs3} (1L  \ch{CaAs3}) at $\Lambda$ is composed of 15.0\% Ba~$5d$ (16\% Ca $3d$) orbitals while the $\Gamma$ valley is composed of only 6\% Ba $5d$ (7\% Ca $3d$). Because the $p$-$d$ interaction is of bonding nature, we can expect the state at $\Lambda$ to be stabilized, i.e., lower in energy with increasing $d$ orbital contribution. However, this alone does not provide an explanation for why the band gap is indirect for 1L  \ch{BaAs3} but not for 1L \ch{CaAs3}.

The indirect band gap of 1L \ch{BaAs3} can be attributed to an $increase$ in the strength of the $p$-$d$ interaction from 1L \ch{CaAs3} to 1L \ch{BaAs3}: The Ba~$5d$ orbital is more delocalized than the Ca~$3d$ orbital because the electrons in the $5d$ orbital experience stronger screening from the nuclear charge. Thus, the Ba~$5d$ states can hybridise more strongly with As~$4p$ orbitals than Ca $3d$. As a result, the stronger bonding $d$-$p$ interactions in 1L \ch{BaAs3} stabilize the $\Lambda$ valley CB state more than the weaker interaction in 1L \ch{CaAs3}. This more pronounced energy shift leads to an indirect band gap in 1L \ch{BaAs3}. Checking the conduction band dispersion of a hypothetical 1L \ch{SrAs3} (data not shown), we find a trend of the $\Lambda$ valley energy decreasing with respect to $\Gamma$ from the $n=3$ to the $n=5$ shell, i.e. from Ca~$3d$ over Sr~$4d$ to Ba~$5d$, which strengthens the $p$-$d$ interaction argument.

Next, we compare the band structures of monolayer \ch{CaAs3} and \ch{CaP3} (Fig.~\ref{Fig:band_DOS}). The conduction band of 1L \ch{CaP3} shows a stronger dispersion than that of 1L \ch{CaAs3}, which is consistent with the prediction of high electron mobilities in 1L \ch{CaP3}\citep{lu_cap3_2018}. 1L \ch{CaP3} has a larger band gap than \ch{CaAs3}. Interestingly, 1L \ch{CaP3} does not show any local CBM at $\Lambda$ between $\Gamma$ and $S$. And this despite the fact that the Ca~$3d$ orbitals contribute about 53 \% of the CB state at $\Lambda$, which is significantly more than for 1L \ch{CaAs3}. We attribute this difference in orbital composition and electronic structure to the increased asymmetry of the 1L \ch{CaP3} structure. The lattice parameter $a = 5.71$~\AA ~is larger than $b = 5.56$~\AA, whereas for 1L \ch{CaAs3} the parameters are almost identical with $a \approx b \approx 5.97$~\AA. This distortion leads to considerable differences in the projected charge density at the band edges for 1L \ch{CaP3} (see supplementary information, Fig. S5).

Finally, we would like to note a similarity between the electronic structure of 1L \ch{CaP3}, \ch{CaAs3}, \ch{BaAs3} and that of hybrid halide perovskites. Specifically, both the top of the valence band and bottom of the conduction band are dominated by \textit{antibonding} orbitals~\citep{umebayashi_electronic_2003, tao_absolute_2019}. Defect tolerance properties inherent to hybrid halide perovskites are attributed to this feature~\citep{kim_role_2014, yin_unusual_2014, zheng_electronic_2019}. Thus, we can anticipate 1L \ch{CaP3}, \ch{CaAs3} and \ch{BaAs3} to exhibit a similar tolerance to native defects since states associated with dangling bonds are expected to appear within the bulk of valence or conduction band states rather than in the band gap. 

\section{Conclusion}
In this work, we calculated the band alignment of puckered \ch{CaP3}, \ch{CaAs3} and \ch{BaAs3} monolayer at the G$_0$W$_0$ level of theory for the isolated monolayers according to the electron affinity rule.
Our calculations suggest that monolayer \ch{CaP3}, \ch{CaAs3} and \ch{BaAs3} all form type-II (staggered) heterojunctions. Their quasiparticle gaps are 2.1 (direct), 1.8 (direct) and 1.5~eV (indirect), respectively. The differences in alignment with respect to the vacuum potential (i.e. the conduction band offsets) result from the different degree of relative stretching of the As or P anionic mesh upon insertion of the Ca or Ba ions.

We also discussed trends in the electronic structure in the light of chemical bonding analysis. We found that the indirect band gap in \ch{BaAs3} is caused by relatively strong As~$3p$ - Ba~$5d$ bonding interactions that stabilize the conduction band at $\Lambda$. 

\clearpage
\begin{suppinfo}

The following files are available free of charge.
\begin{itemize}
  \item VASP structure files (POSCAR) of monolayer \ch{CaP3}, \ch{CaAs3} and \ch{BaAs3}.
  \item Band structures plotted from Wannier functions versus band structures along a k-path from DFT (Fig. S1-S3). These plots allow to assess the accuracy of the interpolation.
  \item Projected charge density of monolayer \ch{CaAs3} at a different isosurface level (0.0025~e/\AA$^3$) (Fig. S4), and projected charge density of monolayer \ch{CaP3} (Fig. S5).
  \item COHP analysis for buckled arsenene and phosphorene (Fig. S6-S7). Clear antibonding nature of $p$-$p$ interactions can be seen at the band edges.
\end{itemize}

\end{suppinfo}
\begin{acknowledgement}

The authors acknowledge funding provided by the Natural Sciences and Engineering Research Council of Canada under the Discovery Grant Programs RGPIN-2020-04788.
Calculations were performed using the Compute Canada infrastructure supported by the Canada Foundation for Innovation under John R. Evans Leaders Fund.

\end{acknowledgement}

\section*{Author contributions}
M.L. conceived the idea in discussions with O.R. and H.S. and performed the calculations and data analysis. H.S. helped with the preparation of the monolayer structure files. The manuscript was written by M.L., revised by O.R. and reviewed by H.S.. The project was supervised by O.R..


\bibliographystyle{manuscript}
\bibliography{manuscript}


\end{document}